# Modeling the spread of the Zika virus using topological data analysis

Derek Lo[1,2,*] and Briton Park[1,3,*]
[1] Department of Statistics, Yale University
[2] Department of Computer Science, Yale University
[3] Department of Mathematics, Yale University
[*] Authors are co-first authors listed in alphabetical order by last name

**Abstract**

Zika virus (ZIKV), a disease spread primarily through the *Aedes aegypti* mosquito, was identified in Brazil in 2015 and was declared a global health emergency by the World Health Organization (WHO). Epidemiologists often use common state-level attributes such as population density and temperature to determine the spread of disease. By applying techniques from topological data analysis, we believe that epidemiologists will be able to better predict how ZIKV will spread. We use the Vietoris-Rips filtration on high-density mosquito locations in Brazil to create simplicial complexes, from which we extract homology group generators. Previously epidemiologists have not relied on topological data analysis to model disease spread. Evaluating our model on ZIKV case data in the states of Brazil demonstrates the value of these techniques for the improved assessment of vector-borne diseases.

**Introduction**

An explosive outbreak of ZIKV began in Brazil in April 2015. The ZIKV outbreak has spread all throughout South America due to the abundance of the *Aedes aegypti* mosquito species[1], which is the primary transmission vector for ZIKV[2]. Approximately 2.6 billion people live in regions suitable for the virus to spread[3], and as of November 2016, 57 countries have active local ZIKV transmission[4]. Researchers project that the number of ZIKV cases in Brazil will be at least double that of any other country[5]. We focus our analysis on Brazil because of the country's unique susceptibility to the virus. The immediate symptoms of ZIKV are mostly mild, but ZIKV has been associated with more serious conditions. ZIKV has been linked to increased cases of the Guillain-Barré Syndrome[6], which is a severe neurological disease that causes the immune system to attack the nervous system. Occurrences of microcephaly, a condition causing the brain to be underdeveloped, has increased in the children of infected pregnant women[7]. Predicting the spread of ZIKV is, therefore, a priority.

Researchers have made substantial progress in modeling ZIKV to better understand and prevent its spread. Various models have already been developed to study the spread of vector-borne diseases. For example, variants of the SIR model from classic epidemiology theory can be used for these purposes and often are applied to obtain estimates of the basic reproduction number, a metric to study how infectious a disease is[8]. For example, Gao *et al.* (2016) used a type of SEIR model based on classic epidemic theory to estimate the impact of mosquito-borne and sexual transmission of the ZIKV in Brazil, Colombia, and El Salvador and estimate a basic reproduction number of the virus[9]. However, this framework requires reliable estimates of epidemiological parameters such as the vector disease transmission rate.

Another class of models that are often used by researchers to study the spread of vector-borne diseases are time series models such as ARIMA[10]. These models can take advantage of time series data of climate conditions and disease incidence data to model disease transmission over time. Zhang *et al.* (2016) developed a time series predictive model for the dengue disease in China using weather predictions and dengue surveillance information in order to obtain projections of the number of ZIKV infections in the Americas[11]. One drawback of time series models is that acquiring granular time-series data over extended durations can be difficult[11].

Researchers have also used stochastic models to analyze ZIKV spread and prevention methods. Castro *et al.* (2016) developed a stochastic model to capture the uncertainty in ZIKV reporting, importation, and transmission to identify regions of greatest risk in the state of Texas[12]. But again, this model again requires estimates of epidemiological parameters like reporting rates and vector abundance[12].

We propose studying the topological features of vector locations to inform the parameter estimation of these models. To validate the explanatory power of these features, we use a linear regression model to predict the number of ZIKV cases in each state of Brazil using features generated from the mosquito population topology. We find that our prediction results are comparable to the performance of models in the existing literature.

**Methods**

Predictions of the number of Zika cases can be obtained by utilizing the population density of *Aedes aegypti* mosquitoes[13], average temperature of a given region[14], and human population density. However, by applying techniques from persistent homology via Vietoris-Rips filtrations—we find valuable information within the spatial structure of the locations of *Aedes aegypti* mosquitoes.

First, we obtain data on the geographic locations of *Aedes aegypti* mosquitos in Brazil. In 2013, Brazilian municipalities conducted physical household surveys, searching for mosquito larvae, pupae, and adult mosquitos. If a mosquito population was discovered in the municipality over a region greater than 5km x 5km, a "mosquito occurrence" was marked as a polygon region[15]. The coordinate associated with this polygon its centroid. There are 5057 entries, each of which has an associated polygon centroid that represents a mosquito population at that location during 2013[16]. An example of our data for Brazil and the state of Sergipe is given in Fig 1. A more detailed description of the data can be found in our supplementary information. To estimate the relationship between the number of ZIKV cases and the spatial structure of the *Aedes aegypti* mosquito populations within each state, we use monthly reports containing the cumulative number of confirmed ZIKV cases in each state, which Brazil's Ministry of Health has been publishing since 2015.

We propose a model that exploits the spatial information within the *Aedes aegypti* mosquito occurrence maps. This is done by utilizing ideas from persistent homology to extract topological information from the 2-dimensional point clouds resulting from the longitudinal and latitudinal coordinates of the *Aedes aegypti* mosquito polygon centroids. Specifically, we study the $0^{th}$ and $1^{st}$ dimensional homology group generators (connected components and loops), which cannot be accessed using more standard statistical techniques. We use the Vietoris-Rips filtration[17] to create intermediate structures (simplicial complexes) from which we can extract topological information from our original data. The Vietoris-Rips filtration is applied to the coordinates of the polygon centroids in each state of Brazil using the TDA package in R[18]. The filtration is constructed starting with balls of radius 0 around each point. As we grow the ε-balls, some begin to intersect with one another, which forms "simplexes". A 0-simplex is defined to be a single vertex, a 1-simplex is a line segment connecting a pair of vertices and a 2-simplex is a triangle connecting three vertices. For each value of ε, we obtain a simplicial complex which is composed of all the simplexes in the filtration. Note that in the filtration, the simplicial complex for a specified value of ε is a subset of the simplicial complexes of larger ε. For further exposition on this filtration process, see supplementary information 1.1 and 1.2.

Using these topological features of the *Aedes aegypti* mosquito occurrence locations in each state of Brazil and the corresponding number of ZIKV cases in each state, we fit linear regression models. More specifically, we look at the number of H0 features at the start of the filtration (H0N), the total number of H1 features throughout the filtration (H1N), and the maximum lifetime of H1 features (H1ML). The H0N features measure the density of mosquitoes, with a higher number of H0 features indicating a greater presence of mosquitos. The H1 features measure the spatial distribution of mosquitoes, with more H1 features indicating loops or areas in the state without mosquito occurrences.

**Results and Discussion**

Researchers have found the effects of mosquito populations and climate to be useful in anticipating the transmission of vector-borne diseases[14,19]. Some have even proposed a fast way of predicting the spread of ZIKV infections in Brazil[5,13], which involves estimating the number of ZIKV cases using attributes such as the average temperature, the population density, and the population of *Aedes aegypti* mosquitos of a region. We use the number of *Aedes aegypti* mosquito occurrences (AMO) in the 2013 physical household surveys in Brazil described above as a proxy for the *Aedes aegypti* mosquito population for each state of Brazil. For the temperature of each state, we use the mean yearly temperature calculated by averaging monthly temperatures in degrees Celsius of 98 weather stations in Brazil in 2010 (TEMP)[20]. To estimate the population density, we divide the estimated resident population size of each state in 2014 by the geographic area of each state in km$^2$ (HPOP)[21].

We propose a model that takes advantage of the spatial information within *Aedes aegypti* mosquito occurrences maps in addition to state-level attributes such as population density and average temperature. We include the number of H0 features to predict the number of ZIKV cases as a topological feature, since states with fewer H0 features have fewer municipalities that have had occurrences of the *Aedes aegypti* mosquito than states with less H0 features. A low number of H0 features for a state could either arise through a fewer number of municipalities or the existence of municipalities that do not harbor *Aedes aegypti* mosquitos. Because ZIKV is mainly transmitted to people from *Aedes aegypti* mosquitoes, either case will lead to fewer infections. A fewer number of municipalities is likely to indicate a lower population density in a state, which would decrease the chance of ZIKV infections, while the existence of municipalities that do not harbor many *Aedes aeypgti* mosquitoes also can lead to a lower chance of ZIKV infections for a state due to a fewer number of transmission vectors. Therefore, we anticipate that ZIKV transmission to people is more likely in the states with more occurrences of the *Aedes aegypti* mosquito or H0 features and less likely in states with fewer occurrences of the mosquito.

We also include the number of H1 features to model the spread of ZIKV. A large number of H1 features within a state may arise due to a large amount of municipalities with *Aedes aegypti* mosquitoes exist within a state, which create more loops during the Vietoris-Rips filtration due to chance or municipalities in general are more spread out due to geographic barriers such as mountains or lakes. Keeping the number of municipalities with *Aedes aegypti* mosquitoes constant within a state, we anticipate that more H1 features will signify that municipalities with the mosquito occurrences are spread out. This spatial distribution of these municipalities may indicate either a low *Aedes aegypti* mosquito density or human population density in a state.

Therefore, we anticipate that a large number of H1 features will have a negative effect on the number of ZIKV infections, keeping the number of H0 features constant. It is worth noting that the H1 features respect geographical topology such as lakes and mountainous regions, since municipalities are far less likely to be located in these areas. Thus it is very improbable to have polygon centroids located in such regions and therefore the loops do not intersect a state's topography.

Lastly, we include H1ML in our model because it informs the disease's transmission rate between municipalities. States with densely packed municipalities will see higher transmission rates due to greater mobility. And conversely, states with municipalities spread far apart due to geographic topology such as mountainous regions and lakes will observe lower transmission rates, and therefore fewer cases. A high H1ML indicates that the municipalities within that state are spread farther apart than states with low H1ML. Furthermore, simply calculating the density of municipalities is not sufficient; the state could be very large and have a few number of closely packed municipalities. Thus, we need some way of determining the proximity of the municipalities between each other. The H1ML feature gives us a proxy for this. We specifically study the maximum lifetime of H1 features for each state rather than an average or median, since this metric is robust to the H1 features that have very short lifetimes which may arise due to a large number of H0 features out of chance rather than being a signal for areas free of *Aedes aegypti* mosquitos.

Using these features, we create a linear regression model. This model uses AMO (H0N), TEMP, POP, H1N, H1ML as its predictors and the number of ZIKV cases with a logarithmic transformation as the response. The model is labeled A (Table 1). All of the model's features, except POP and H1ML, are statistically significant at the 5% level, while POP is significant at the 10% level. AMO (H0N), POP, and TEMP have positive effects on the number of ZIKV cases, while the interaction between H1ML and H1N have negative effects. To compute the fit of the model, we calculate the model's adjusted $R^2$ and find it to be 0.64. Additionally, to test if a linear regression model is appropriate for the data, we look at the residuals of the model (Figure 1). We find that the residuals are approximately normally distributed, which imply that a linear regression model is appropriate. To test how well the model predictions perform, we use leave-$p$-out cross-validation and achieve average errors of 0.85, 0.41, and 0.91 using $p$ = 1, 2, and 3, respectively.

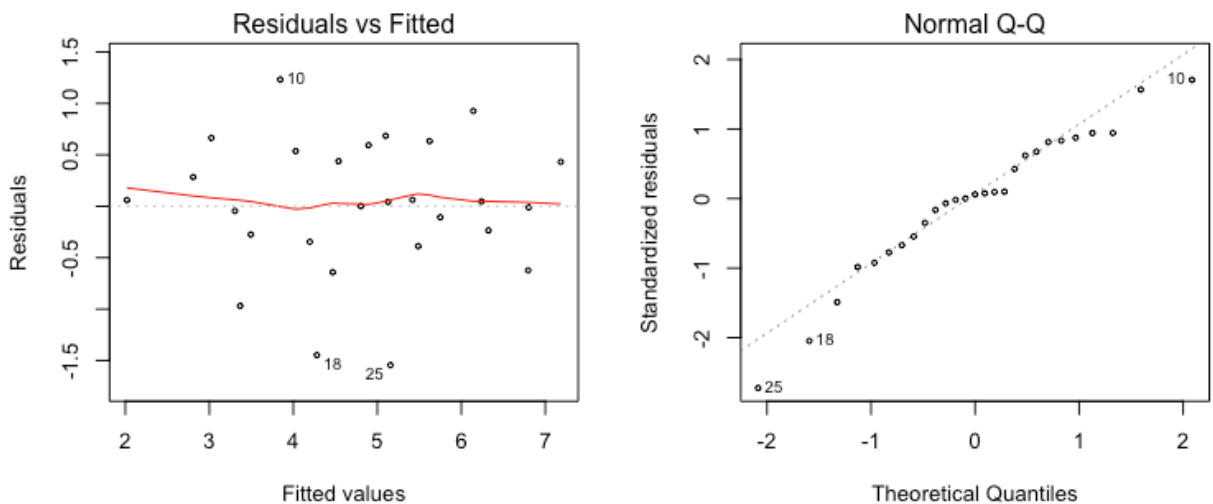

**Fig 1. Fitted vs Residuals Plot and Q-Q Plot of Standardized Residuals of Model A.** The two plots show that the residuals seem to be symmetrically distributed around 0 and in general clear patterns are not present. Therefore, the residuals are approximately normally distributed around 0 and a linear regression with a logarithmic transformation of the number of ZIKV cases appropriately models our data.

**Table 1. Coefficients of linear regression model predictors**

| Model | Model A | Model B |
|---|---|---|
| Intercept | -3.79* | -3.77 |
| AMO / H0N | 0.059* | 0.0067* |
| H1N | -0.11* | - |
| H1ML | 0.0052 | - |
| Interaction (H1N and H1ML) | -0.090* | - |
| POP | 0.0057* | 0.0060* |
| TEMP | 0.22* | 0.27* |

*Coefficient is statistically significant at the 5% significance level

**Table 2. Leave-*p*-out cross-validation mean squared errors**

| p | Model A | Model B |
|---|---|---|
| 1 | 0.85 | 1.91 |
| 2 | 0.43 | 1.99 |
| 3 | 0.91 | 2.07 |

It is possible that this final model overfits the data given its higher number of topological predictors. To check for this potential problem, we fit another linear regression model by removing the topological features predictors; this model is labeled B in Table 1. The model's goodness-of-fit can be measured through the adjusted $R^2$, which was 0.41. We also apply leave-$p$-out cross-validation[22] using $p$ = 1, 2, and 3 to test how well the model predicts the log-transformed number of ZIKV cases in each state of Brazil. The model achieves an average cross-validated squared error of 1.91, 1.99, and 2.07 for $p$ = 1, 2, and 3, respectively (Table 2). Therefore, we see that model A is an improvement over the reduced model B when evaluating them on the prediction errors and the adjusted $R^2$ metric.

We compare the results of our model with the results of Zhang *et al.* (2016). Zhang *et al.* (2016) report a Pearson correlation value of 0.57 between state-level model projections of ZIKV cases and surveillance data of ZIKV cases through June 2016 in Colombia. We obtain a Pearson correlation value of 0.71 between state-level model projections of ZIKV cases and the cumulative number of actual ZIKV cases through July 2016 in Brazil and plot the predicted number of cases against the confirmed ZIKV cases (Figure 2). Due to our high correlation value, our model performance is comparable to the results in Zhang *et al.* (2016). Thus, we show the explanatory power of the integration of state-level attributes and topological features in predicting the number of ZIKV cases through using even simple linear regression.

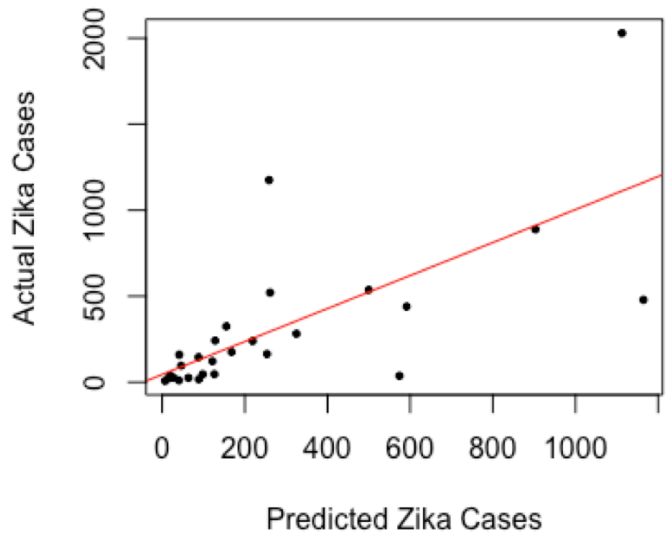

**Fig 2. Predicted Versus Actual Confirmed Zika Cases for July 2, 2016.** We plot the predicted ZIKV cases against the confirmed Zika cases. The correlation between the predictions and the confirmed cases is 0.71.

Overall, we have shown that topological features of the locations of mosquito occurrences contain additional information that can be used in conjunction with standard features to better predict the spread of ZIKV. Due to the nature of vector-borne diseases, infected arthropod species are their primary modes of transmission. Our results suggest that applying TDA to their locations can help epidemiologists and public-health officials better track vector-borne diseases and curb the spread of future contagions.

**Acknowledgments**
We thank J. Cisewski, A. Lo and M. Lo for comments.

# Supplementary Information

**Table of Contents**

# Methods

## 1. Topological Data Analysis

Topological data analysis (TDA) is a growing field of statistics that makes use of techniques from topology to analyze data. We believe TDA provides valuable tools that epidemiologists can use to improve modeling the spread of disease. Specifically, we will explain the very basics of persistent homology, such as the Vietoris-Rips filtration, as well as how to carry out the filtration using the R programming language. Finally, we explain the persistence diagram, which is a common way of visualizing and extracting information from the Vietoris-Rips filtration.

1.1 Persistent Homology

The central idea of persistent homology is to compute topological features of data through a series of polyhedra, which are simplicial complexes based on a parameter, $\varepsilon$. First to understand what a simplicial complex is, we define the notion of a simplex. In geometry, a simplex is a generalization of a triangle or tetrahedron to n dimensions. For example, a k-simplex is a k-dimensional convex hull composed of k + 1 vertices (a 2-simplex is a triangle). A simplicial complex is then a combination of simplices. Each simplicial complex is contained in the subsequent simplicial complex with a larger $\varepsilon$ parameter. After obtaining a simplicial complex, its homology vector space can be computed. Each element of the homology vector space represents a type of structure in the complex. But does this structure also exist in the data? To answer this question, we track each homology element (or homology class) as the $\varepsilon$ parameter grows. If the feature is statistical noise, the existence of that homology class will likely be short. If the feature is a real signal, the homology class is likely to persist for a longer time[1]. Hence, this method is called *persistent* homology.

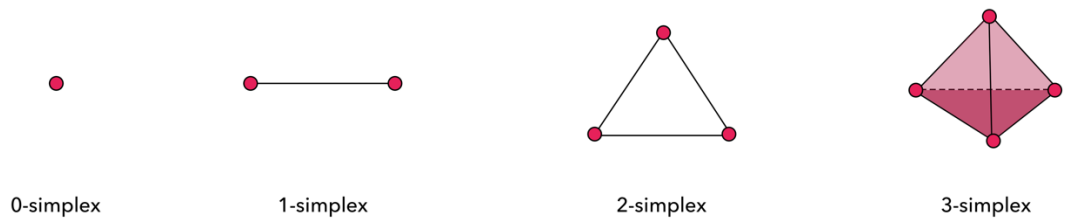

Supplementary Figure 1: An example of simplexes from dimension 0 to dimension 3. These are used to build a simplicial complex, which is the product of a Vietoris-Rips Filtration. See section 1.2 for a detailed walkthrough of this filtration.

1.2 Introduction to the Vietoris-Rips Filtration

The Vietoris-Rips filtration is a popular tool in topological data analysis because it can be used to encode valuable information about the underlying topology of data. The filtration produces a set of simplicial complexes that add a topological skeleton to point clouds. However, it is good to note that in practice, datasets are often too large for the filtration to be calculated in a computationally feasible manner.

To build a complex from a point cloud data set using the Vietoris Rips filtration, we want to approximate the data by a polyhedron, which is a simplicial complex. We define the simplex inductively. The 0-simplices, the vertices of the complex, are defined by the data points. We first initialize a small parameter ε starting at an ε of 0. We surround each vertex with a 2D ball with radius equal to ε and grow these ε balls. If two ε balls intersect, we create a 1-simplex by joining the two 0-simplices at the center of these balls to create a line segment. If 3 ε balls intersect, we create a 2-simplex by joining the three 0-simplices at the center of the 3 balls in a triangle. Now we construct the homology vector spaces by finding the cycles in the complex which are the generators of the homology vector space. A cycle is a simplex which is *not* part of the boundary of a simplex of higher dimension. For example, the vertices of a 1-simplex are not cycles. The components of the polyhedron are the 0-cycles. A loop of 1-simplices, which is not the boundary of a 2-simplex (triangles in the complex) is a 1-cycle. So the 0-cycles count the components of the polyhedron and the 1-cycles count the non-trivial loops in the polyhedron. Basically the cycles are just the holes of various dimension in the complex. These cycles are the generators of the homology vector spaces denoted by H0, H1, H2 for each dimension. The dimension of the vector spaces H0N, H1N are the indices we compute and interpret for the mosquito data.

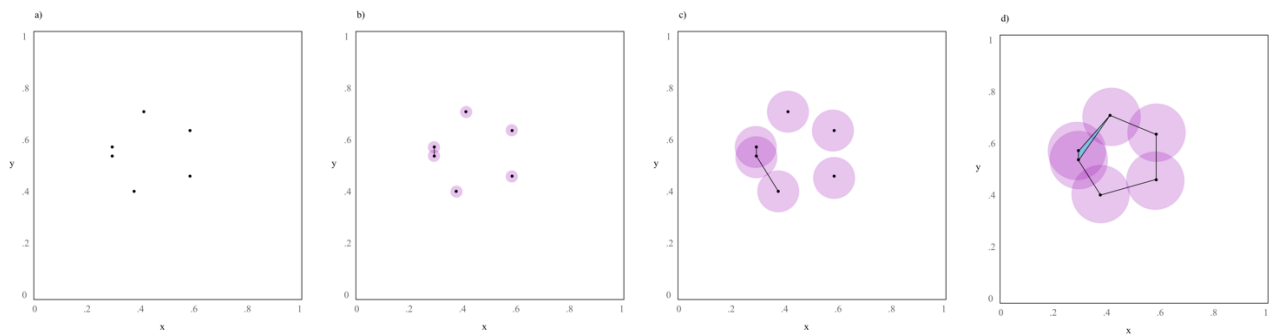

Supplementary Figure 2: A visualization at various ε values of the Vietoris-Rips filtration applied to a toy dataset. Notice how in panel a, we begin with ε = 0 since there are no balls present, and we have six 0-simplices. In panel b we begin to grow ε, and we observe that two points become connected to each other, forming our first 1-simplex. In panel c as we grow ε even further, another pair of points becomes connected, birthing another 1-simplex. Finally, in panel d ε has reached a large enough value that an open loop forms, as well as a closed loop in blue.

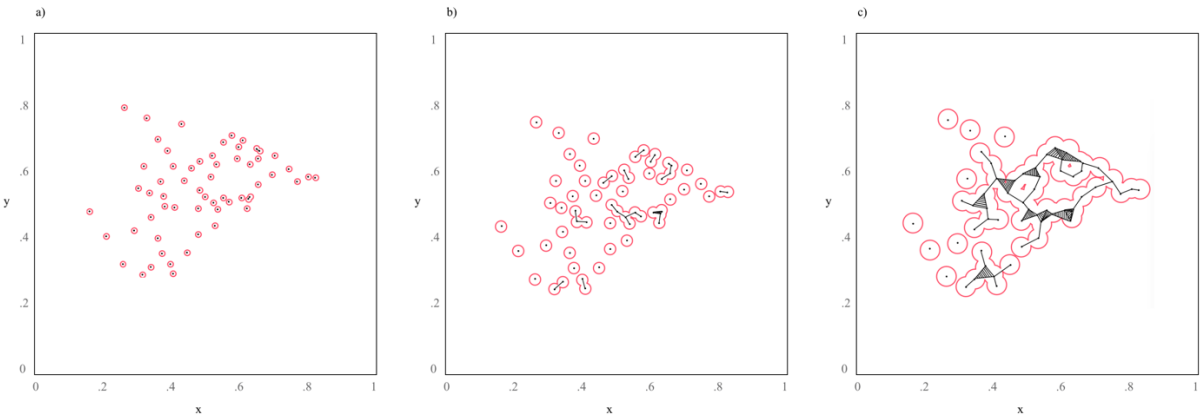

Supplementary Figure 3: Snapshots of the Vietoris-Rips filtration applied to Aedes aegypti mosquito population polygon centroid locations in Sergipe at several ε values. a-c, Depictions of the Vietoris-Rips filtration on the Sergipe mosquito map for $\varepsilon = 0.1$, 0.2, and 0.5. Black dots represent 0-simplexes, lines represent 1-simplexes, and shaded triangles represent 2-simplexes, while red circles denote $\varepsilon$-balls. 2-simplexes are filled because they are, by definition, closed loops.

There are several existing software packages that can compute the Vietoris-Rips filtration, including the TDA package in R, Perseus, and Dionysus.

1.3 R code for Vietoris-Rips Filtration

```
library(TDA)

Sao_Paulo <- ripsDiag(states[states$V13 == "Sao Paulo",c(6,7)], 1, 3, dist = "euclidean")
plot(Sao_Paulo$diagram, main = "Sao_Paulo, 479", band = 2*cc1_0)

Maranhao <- ripsDiag(states[states$V13 == "Maranhao",c(6,7)], 1, 2, dist = "euclidean")
plot(Maranhao$diagram, main = "Maranhao, 282")

Santa_Catarina <- ripsDiag(states[states$V13 == "Santa Catarina",c(6,7)], 1, 20, dist =
"euclidean")
plot(Santa_Catarina$diagram, main = "Santa Catarina, 8")

Amapa <- ripsDiag(states[states$V13 == "Amap\xcc\xc1",c(6,7)], 1, 20, dist = "euclidean")
plot(Amapa$diagram, main = "Amapa, 11")

Acre <- ripsDiag(states[states$V13 == "Acre",c(6,7)], 1, 5, dist = "euclidean")
plot(Acre$diagram, main = "Acre, 40")

Alagoas <- ripsDiag(states[states$V13 == "Alagoas",c(6,7)], 1, 0.8, dist = "euclidean")
plot(Alagoas$diagram, main = "Mato Alagoas, 325")

Amazonas <- ripsDiag(states[states$V13 == "Amazonas" | states$V13 == "State of Amazonas",c(6,7)],
1, 10, dist = "euclidean")
plot(Amazonas$diagram, main = "Amazonas, 25")

Bahia <- ripsDiag(states[states$V13 == "Bahia",c(6,7)], 1, 2, dist = "euclidean")
plot(Bahia$diagram, main = "Bahia, 1175")
```

```r
Ceara <- ripsDiag(states[states$V13 == "Cear\xcc\xc1",c(6,7)], 1, 1, dist = "euclidean")
plot(Ceara$diagram, main = "Ceara, 521")

Espirito_Santo <- ripsDiag(states[states$V13 == "Esp\xcc_rito Santo",c(6,7)], 1, 1, dist =
"euclidean")
plot(Espirito_Santo$diagram, main = "Espirito_Santo, 160")

Gois <- ripsDiag(states[states$V13 == "Goi\xcc\xc1s",c(6,7)], 1, 1.5, dist = "euclidean")
plot(Gois$diagram, main = "Gois, 145")

Mato_Grosso <- ripsDiag(states[states$V13 == "Mato Grosso",c(6,7)], 1, 2, dist = "euclidean")
plot(Mato_Grosso$diagram, main = "240")

Mato_Grosso_Sul <- ripsDiag(states[states$V13 == "Mato Grosso do Sul",c(6,7)], 1, 3, dist =
"euclidean")
plot(Mato_Grosso_Sul$diagram, main = "Mato Grosso Sul, 22")

Minas_Gerais <- ripsDiag(states[states$V13 == "Minas Gerais",c(6,7)], 1, 2, dist = "euclidean")
plot(Minas_Gerais$diagram, main = "Minas Gerais, 122")

Para <- ripsDiag(states[states$V13 == "Par\xcc\xc1",c(6,7)], 1, 3, dist = "euclidean")
plot(Para$diagram, main = "Para, 46")

Paraiba <- ripsDiag(states[states$V13 == "Para\xcc_ba",c(6,7)], 1, 0.6, dist = "euclidean")
plot(Paraiba$diagram, main = "Paraiba, 889")

Parana <- ripsDiag(states[states$V13 == "Paran\xcc\xc1",c(6,7)], 1, 1.5, dist = "euclidean")
plot(Parana$diagram, main = "Parana, 37")

Pernam <- states[states$V13 == "Pernambuco",c(6,7)]
Pernam <- Pernam[-(nrow(Pernam)-1),]
Pernambuco <- ripsDiag(Pernam, 1, 6, dist = "euclidean")
plot(Pernambuco$diagram, main = "Pernambuco, 2029")

Piaui <- ripsDiag(states[states$V13 == "Piau\xcc_",c(6,7)], 1, 2, dist = "euclidean")
plot(Piaui$diagram, main = "Piaui, 176")

Rio_de_Janeiro <- ripsDiag(states[states$V13 == "Rio de Janeiro",c(6,7)], 1,1, dist =
"euclidean")
plot(Rio_de_Janeiro$diagram, main = "Rio de Janeiro, 537")

Rio_Grande_do_Norte <- ripsDiag(states[states$V13 == "Rio Grande do Norte",c(6,7)], 1, 1.2, dist
= "euclidean")
plot(Rio_Grande_do_Norte$diagram, main = "440")

Rio_Grande_do_Sul <- ripsDiag(states[states$V13 == "Rio Grande do Sul",c(6,7)], 1, 3, dist =
"euclidean")
plot(Rio_Grande_do_Sul$diagram, main = "Rio Grande do Sul, 96")

Rondnia <- ripsDiag(states[states$V13 == "Rondnia",c(6,7)],  2, 2, dist = "euclidean")
plot(Rondnia$diagram, main = "Rondnia, 17")

Roraima <- ripsDiag(states[states$V13 == "Roraima",c(6,7)], 1, 2, dist = "euclidean")
plot(Roraima$diagram, main = "Roraima, 26")

Sergipe <- ripsDiag(states[states$V13 == "Sergipe",c(6,7)], 1, 1, dist = "euclidean")
plot(Sergipe$diagram, main = "Sergipe, 242")

Tocantins <- ripsDiag(states[states$V13 == "Tocantins",c(6,7)], 1, 1.5, dist = "euclidean")
```

```
plot(Tocantins$diagram, main = "Tocantins, 164")
```

1.4 Persistence Diagram

Typically, persistent homology is visualized through either a barcode diagram or a persistence diagram. A persistence diagram is a plot of the birth and death times of each topological feature in every dimension. The first coordinate of each point denotes the birth time of the feature, while the second coordinate denotes the death time. Formally, a persistence diagram is a set of points expressed as: $\{(u, v) \in R^2 \mid u, v \geq 0, u \leq v\}$. Therefore, all points lie on or above the line of equality. Points close to the line of equality denote features with shorter lifetimes, while points that lie farther away from the line denote features that persist for a longer time.

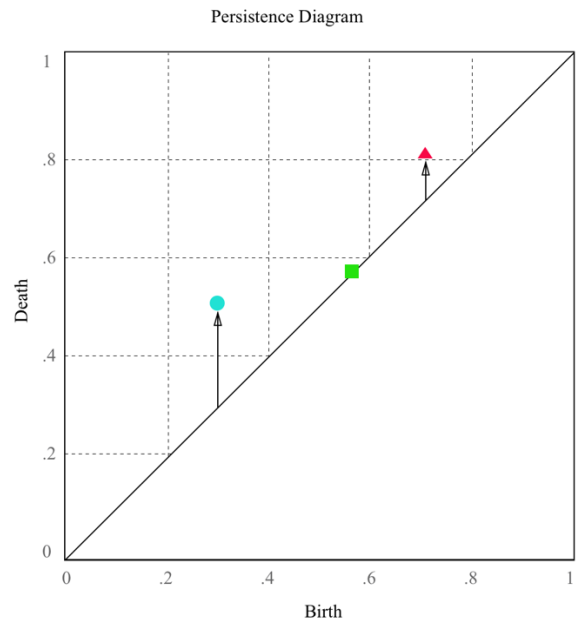

Supplementary Figure 4: A persistence diagram that summarizes information from a Vietoris-Rips filtration. The lengths of the vertical arrows represent the lifetime of the respective features that they point to. The green square is a feature that was born and died at the same time. This can happen when a loop is formed but is immediately closed at once.

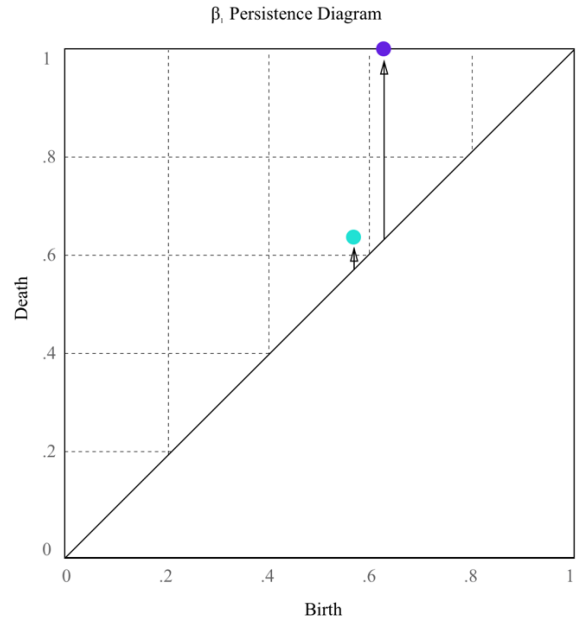

Supplementary Figure 5: Persistence Diagram that depicts specifically the 1D topological features of the example filtration from 1.2, visualized in Supplementary Figure 1. Notice that the blue circle has a very short lifetime, and it corresponds to the blue closed triangle in panel d of Supplementary Figure 1. In that same panel, we notice that a large outer loop has formed with all of the points. This corresponds to the purple point, which has a death time of 1 since the feature lives past the end of the filtration.

**2. Statistical Analysis**

2.1 Modeling using topological features

We are interested in studying the 0-dimensional homology group generators (H0 features) and 1-dimensional homology group generators (H1 features) of the simplicial complexes representing the original point clouds of mosquito occurrences of each state of Brazil. The number of H0 features at the the beginning of the filtration correspond to the number of *Aedes aegypti* mosquito occurrences in each state, since each occurrence starts out as a separate connected component as 0-simplexes in the filtration. Therefore, the H0 features all have birth time of $\varepsilon = 0$. As we grow $\varepsilon$, 1-simplexes are created from intersections of the $\varepsilon$-balls, and the number of connected components decreases. When a connected component dies by joining another connected component, we note its death time or the $\varepsilon$ value at which the component dies. The H1 features correspond to the empty loops born in the Vietoris-Rips filtration. The H1 features die in the filtration when they are filled in by 2-simplexes. We note the birth and death time of these topological features, which are visually displayed in persistence diagrams. We then compute the lifetimes of the H1 features using the birth and death times in the persistent diagrams and record the maximum lifetime for the H1 features for each filtration.

2.2 R Code for Modeling

## I. Linear Regression

```
lmA <- lm(log(Brazil$Cases_latest) ~  Brazil$Lifetime_1D*Brazil$Number_1D + Brazil$Number_0D
+Brazil$temp + Brazil$Population_density )

lmB <- lm(log(Brazil$Cases_latest) ~ Brazil$Number_0D +  Brazil$Population_density + Brazil$temp)

summary(lmA)
summary(lmB)
AIC(lmA)
BIC(lmB)
```

## II. Cross Validation

```
### Leave-1-out cross validation

error12 <- rep(0,27)
error1 <- rep(0,27)

Brazil$pred1 <- 0
Brazil$pred12 <- 0
for(i in 1:27){
  A <- Brazil[i,]
  B <- Brazil[-i,]
  lm1 <- lm(log(Cases_latest) ~ Number_0D +  Population_density + temp, data = B)
  lm12 <- lm(log(Cases_latest) ~ Lifetime_0D_R*Number_1D + Number_0D  + temp  +
Population_density, data = B)
  Brazil$pred1[i] <-  exp(predict(lm1, A))
  Brazil$pred12[i] <-  exp(predict(lm12, A))
  error1[i] <-  log(A$Cases_latest) - predict(lm1, A)
  error12[i] <- log(A$Cases_latest) - predict(lm12, A)
}

sum(error1**2)/27
sum(error12**2)/27

### Leave-2-out cross validation

error1 <- rep(0,351)
error12 <- rep(0,351)
count <- 0
for(i in 1:26){
  for(j in (i+1):27){
      count <- count + 1

      train <- Brazil[-c(i,j),]
      test <- Brazil[c(i,j),]
      lm1 <- lm(log(Cases_latest) ~ Number_0D +  Population_density + temp, data = train)
      lm12 <- lm(log(Cases_latest) ~Number_1D + Life_time_0D_R + Lifetime_1D*Number_0D + temp +
Population_density, data = B)
      error1[count] <-  mean((log(test$Cases_latest) - predict(lm1, test))**2)
      error12[count] <- mean((log(test$Cases_latest) - predict(lm12, test))**2)
  }
}
sum(error1)/351
sum(error12)/351

### Leave-3-out cross validation
```

```
X <- combn(1:27, 3)
count <- 0
error1 <- rep(0,2925)
error12 <- rep(0,2925)
for(i in 1:2925){
  count <- count + 1
  train <- Brazil[-X[,i],]
  test <- Brazil[X[,i],]
  lm1 <- lm(log(Cases_latest) ~ Number_0D +  Population_density + temp, data = train)
  lm12 <- lm(log(Cases_latest) ~Number_1D + Life_time_0D_R + Lifetime_1D*Number_0D + temp + 
Population_density, data = train)
  
  error1[count] <-  mean((log(test$Cases_latest) - predict(lm1, test))**2)
  error12[count] <- mean((log(test$Cases_latest) - predict(lm12, test))**2)
}
mean(error1)
mean(error12)
```

### 3. Code and Data Availability

*Aedes aegypti* **Mosquitos**

Our mosquito data comes from the Global Compendium of *Aedes aegypti* and *Ae. albopictus* occurrence dataset[2]. The full dataset is hosted by the Dryad Digital Repository at http://dx.doi.org/10.5061/dryad.47v3c. In total we consider 5057 entries for Brazil. Each entry represents a mosquito population in a region called a polygon, which is an area with dimension greater than 5km x 5km. Each polygon represents a survey conducted within a Brazilian municipality with a positive finding of a mosquito population of non-zero size. No attempt was made to quantify the true number of mosquitos within a polygon. There are 5570 municipalities in Brazil.

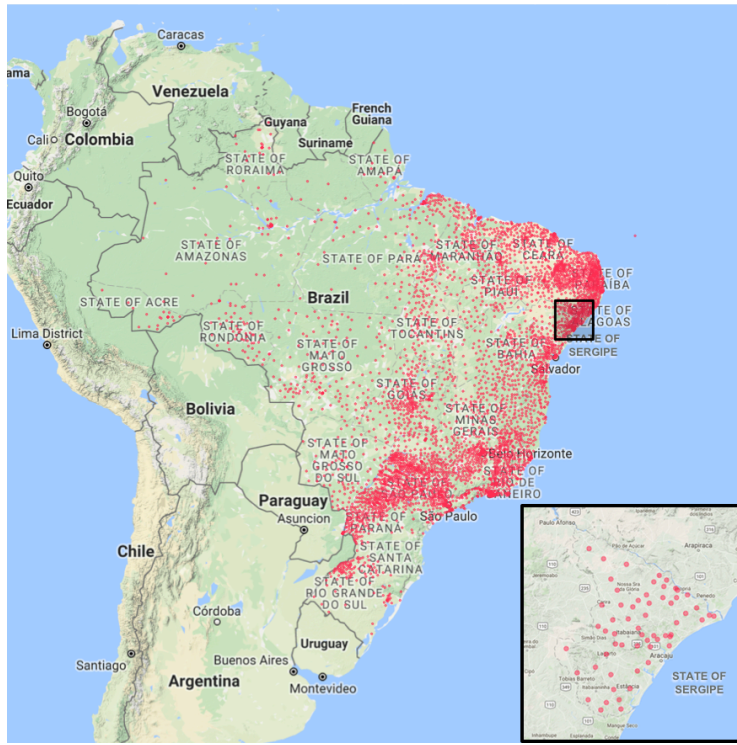
22

Supplementary Figure 6: A map of *Aedes aegypti* mosquito occurrences in Brazil**.** Red dots indicate the location of each mosquito occurrence in 2013. In total there were 5057 mosquito occurrences in Brazil and 60 mosquito occurrences in the state of Sergipe.

**Temperature**

We obtained mean monthly temperatures in degrees Celsius from 98 weather stations which cover all states of Brazil in 2010 using FAOClim-Net, an agroclimatic database management system[3]. The data is available at http://geonetwork3.fao.org/climpag/. We take the average of these monthly temperatures to obtain a mean annual temperature for each state of Brazil in the year 2010. We use these temperatures as a proxy for the temperatures of each state of Brazil in recent years.

**Population Density**

We obtained population data and geographic area (km) for each state of Brazil from the Instituto Brasileiro de Geografia e Estatistica (IBGE)[4]. We use the estimated resident populations in states of Brazil in the year 2014. To obtain an estimated population density for each state, we divide the estimated resident population data by the geographic area (km). The data can be downloaded at http://downloads.ibge.gov.br/downloads_estatisticas.htm

**Zika Cases**

The data on the number of Zika cases in each state of Brazil come from monthly reports published by Brazil's Ministry of Health[5]. Although the Ministry of Health publishes cases that are under investigation, we only used data on the confirmed cases. We used the latest cumulative data published weekly, which is from July 2, 2016. There are 27 states in Brazil, so we have 27 data points. The data can be obtained from the PDF available at:
http://portalsaude.saude.gov.br/index.php/o-ministerio/principal/leia-mais-o-ministerio/197-secretaria-svs/20799-microcefalia

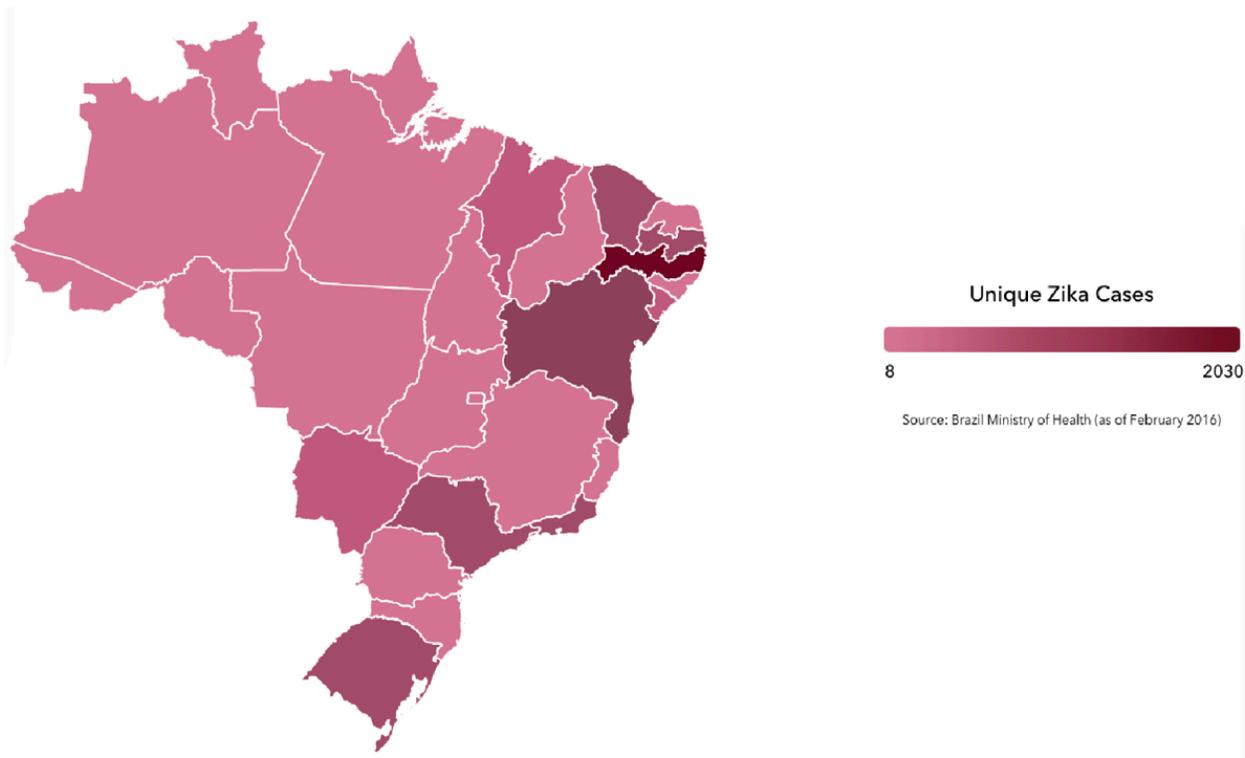

Supplementary Figure 7: A heatmap of *Aedes aegypti* mosquito occurrences in Brazil**.** Red dots indicate the location of each mosquito occurrence in 2013. In total there were 5057 mosquito occurrences in Brazil and 60 mosquito occurrences in the state of Sergipe.

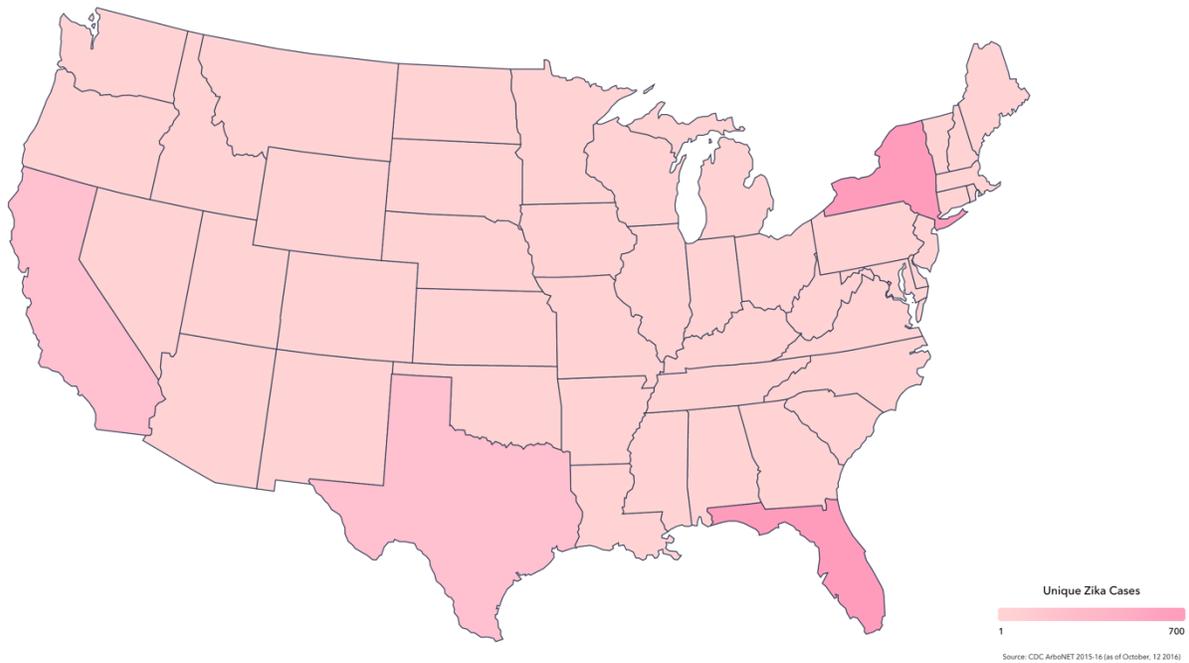

Supplementary Figure 8: A heatmap depicting the unique confirmed Zika cases in the United States. When compared to Supplementary Figure 7, it is clear that there are far more confirmed cases in Brazil than in the United States. The maximum number of Zika cases in the United States is in Florida at 700. By comparison, the maximum number in Brazil is 2030, in Pernambuco.